\documentclass[aps,prl,twocolumn]{revtex4-2}
\usepackage{nameref}
\usepackage[
    colorlinks=true,
    citecolor=blue,
    linkcolor=blue,
    urlcolor=blue
]{hyperref}
\usepackage{amsmath}
\usepackage{amssymb}
\usepackage{caption}
\usepackage{graphicx}
\usepackage{bm}
\usepackage{subcaption}
\usepackage[justification=raggedright,singlelinecheck=false]{caption}
\usepackage{blindtext}
\usepackage[toc,page]{appendix}

\begin{document}

\title{Robust protocols to reveal anyonic time-exchange phase}
\author{In\`es Safi}

\begin{abstract}
We consider hierarchical quantum Hall edge states with $N$ modes and a
spatially local quantum point contact (QPC). In general, the field of an
injected anyon does not directly acquire the universal statistical phase
$\theta$. Short-range inter-edge interactions split the universal anyon
charge and phase into $N$ fractionalized charges associated with
nonuniversal phases $\pi\delta_m$. In contrast, their sum
$\delta=\sum_{m=1}^N\delta_m$, which defines the local scaling dimension
at the QPC, remains protected and is tied to the statistical angle
through
$\pi\delta=\theta$. If the injected anyon is of the same species as the
one dominating backscattering at the QPC, time-domain braiding with
phase $\theta$ is recovered either in the absence of inter-edge
interactions with equal mode velocities, or by performing a spatially
local anyon injection at the QPC.

We then exploit a more robust \emph{local} anyonic time-exchange (ATE)
link between anyons and quasiholes at the QPC, which is also a necessary
ingredient for realizing such braiding. This allows us to propose
minimal single-QPC protocols that do not rely on diluted anyon sources
and that disentangle the role of $\theta$ as a genuine statistical phase
from that as a scaling dimension. From the ATE link we derive two novel
nonequilibrium fluctuation--dissipation relations (FDRs) that isolate
$\theta$. They relate the DC backscattering noise either to an integral
over the DC current or to the phase shift of the AC current with respect
to an applied AC voltage (i.e., the phase of the admittance), accessible
down to low frequencies. For thermalized edges, we show that in the
quantum regime this admittance phase directly yields $\theta$ whenever
$\delta>1/2$.

\end{abstract}

\maketitle
Two-dimensional electronic systems can host quasiparticles that transcend the
boson-fermion dichotomy. A paradigmatic example is the fractional quantum Hall
(FQH) regime, where elementary excitations carry fractional charge $e^*$~\cite{laughlin_fractional_charge_PRL_83,Feldman_review_2021}
and obey unconventional exchange statistics, hence the name \emph{any}(-)ons~\cite{fractional_statistics_anyons_leinaas_myrheim_1977}.
In Abelian FQH states, exchanging two quasiparticles multiplies the many-body wave
function by $e^{i\theta}$, where $\theta$ is the statistical phase. For Laughlin
states at simple filling factors $\nu=1/(2n+1)$, with integer $n$, one has $e^*=\nu e$ and $\theta=\nu\pi$.
More complex filling factors $\nu_C$, such as those of the Jain sequence, remain
only partially understood. They are often addressed within low-energy effective theories that predict several
quasiparticle species with distinct charges and statistical phases~\cite{wen_review_FQHE_1992,*sukho_mac_zhender_PRB_2009}.

Quantum transport through a Hall bar incorporating a quantum point contact (QPC)
has long provided the main experimental access to FQH quasiparticles expected to tunnel through the QPC. Determination of their fractional
charge, a prerequisite for probing statistics, has been extracted from DC shot-noise
measurements obeying a poissonian Fluctuation-Dissipation Relation
(FDR)~\cite{levitov_reznikov}, as validated in a wide variety of experiments~\cite{saminad,*heiblum_frac_97,kane_fisher_noise,*ines_resonance,kyrylo_FQHE_2024}.
This FDR was generalized beyond bipartite systems and inversion symmetry within the
\emph{unifying nonequilibrium perturbative} (UNEP) framework~\cite{ines_degiovanni_2016,ines_PRB_R_noise_2020},
thereby avoiding the need for a detailed microscopic description of FQH edges.

 UNEP theory has also provided \emph{model-independent} protocols to extract the
fractional charge $e^*$ from time-dependent transport
\cite{ines_eugene,*ines_PRB_2019,*ines_photo_noise_PRB_2022,ines_cond_mat,ines_PRB_R_noise_2020}.
These approaches have proved particularly valuable for experiments
performed at complex filling factors $\nu_C$~\cite{glattli_photo_2018,ines_gwendal}.
In practice, deviations from Tomonaga--Luttinger liquid (TLL)
predictions\footnote{It is worth emphasizing that the determination of
fractional charge from photo-assisted current proposed in
Refs.~\cite{wen_photo_PRB_91,crepieux_photo} relies on a specific TLL
framework, which is inconsistent with several experimental
observations. Moreover, these works do not correctly capture the
zero-temperature limit near resonant values of the DC voltage, where
divergences appear-at precisely the regime that is essential for a
reliable charge determination, as demonstrated in
Ref.~\cite{ines_imen_2025}.} are routinely observed. These deviations
may originate from edge reconstruction, inter-edge interactions leading
to charge fractionalization~\cite{ines_epj,fractionnalisation_eugene_mach_zhender_prb_2008,fractionnalisation_review_physics_18_fujisawa},
or from the spatial extension of the QPC. All these effects can be
consistently incorporated within the UNEP formalism.

By contrast, accessing anyonic \emph{statistics} has proven to be substantially more
challenging. Two pioneering DC-based strategies have not been totally successful %
~\cite{fractional_statistics_comment_Read_Nat_Phys_2023,
fractional_statistics_comment_Kivelson_Journal_club_2023}.
The first one relies on interferometric geometries%
~\cite{
fractional_statistics_noise_FPI_FQHE_rosenow_PRB_2012,
manfra_FQHE_statistics_Naturephys_2020,
manfra_2_5_braiding_FPI_FQHE_PRX_2023,
fractional_statistics_anyons_FQHE_heiblum_MZI_2_5_Nature_physics_2023}
, which are sensitive to Coulomb interactions%
~\cite{manfra_braiding_anyons_2022},
thereby complicating the interpretation of the observed phases, especially at $\nu_C$
~\cite{manfra_2_5_braiding_FPI_FQHE_PRX_2023,fractional_statistics_anyons_FQHE_heiblum_MZI_2_5_Nature_physics_2023}.
A second route is based on current cross-correlations, including Hanbury Brown-Twiss
geometries~\cite{ines_prl,*ines_guyon,kim_hbt,*vish_hall_hbt_2003,gefen_HBT_FQHE_PRL_2012,*giuliano_HBT_FQHE_2016}
and QPC-based "anyon colliders" injecting diluted quasiparticle beams~\cite{fractional_statistics_theory_2016,fractional_statistics_theory_Sim_Nat_comm_2016,*Sim_non_abelian_NATcomm_2022,mora2022anyonicexchangebeamsplitter}.
Such colliders have been implemented in Laughlin states~\cite{fractional_statistics_gwendal_science_2020,fractional_statistics_gwendal_PRX_2023,*pierre_anyons_PRX_2023,fractional_statistics_heiblum_sim_nature_2023}, but
cross-correlations alone cannot disentangle $\theta$ from the TLL scaling dimension
$\delta$ \footnote{The scaling dimension $\delta$ is defined from the Green’s function of the quasiparticle field $\Psi$ as $\langle \Psi^{\dagger}(x,t)\Psi(x,0)\rangle \simeq (\omega_c t )^{-\delta}$, where $\omega_c$ denotes a high-energy cutoff. In the zero-temperature limit, the DC current—obtained from the Fourier transform of the correlation functions of the backscattering operator $A$—scales as $(\omega_{\scriptscriptstyle\mathrm dc} / \omega_c)^{2\delta - 1}$, increasing as the scaling dimension $\delta$ decreases. This letter contradicts the widely spread statement that the scaling dimension $\delta$ is not topologically protected agaisnt interactions.
} and significant discrepancies with TLL-based predictions persist at $\nu_C$.~\cite{fractional_statistics_gwendal_PRX_2023,*pierre_anyons_PRX_2023,fractional_statistics_width_anyons_rosenow_PRL_2024,*martin_anyons_finite_width_FQHE_PRL_2024}.

More recently, the experimental analysis of the Hong--Ou--Mandel dip
\cite{feve_anyons_martin_Science_2025},  injecting fractional charges
 shaped by voltage pulses as proposed in
\cite{ines_ann}, has succeeded in disentangling $\theta$ from
$\delta$. However, the underlying theoretical works are based on a TLL behavior for which there is no clear experimental evidence 
\cite{martin_fractional_statistics_dip_HOM_PRL_2023,matteo_PRB_anyons_2025}
and have been revisited in 
\cite{ines_alex_2025}.

In this Letter, we reassess the relation between the time-domain braiding
phase, scaling dimension, and universality by analyzing nonequilibrium
anyonic injection in $N$ co-propagating modes with short-range inter-edge
interactions and a spatially local QPC. We show that the injected anyon
phase forms kinks whose heights are not topologically protected,
yielding $N$ phases associated with the fractionalized charges in each
mode. Only their sum, the local scaling dimension, is robust and obeys
$\delta=\theta/\pi$, which can be accessed through local injection at
the QPC. By contrast, the full phase $\theta$ can be recovered without
additional conditions for \emph{stationary} injection, a result rooted
in current conservation. This resolves the apparent tension between the
nonuniversal renormalization parameter $\lambda$ due to interactions ~\cite{fractional_statistics_theory_2016} and the
identification $\lambda=\theta/\pi$ obtained when these are
ignored~\cite{fractional_statistics_theory_Sim_Nat_comm_2016,
*Sim_non_abelian_NATcomm_2022,oreg_PRL_FQHE_statistics_2023}.
Our analysis therefore restores the robust and \textit{model-independent}
UNEP-based route to extract $\theta$ from the effective voltage drop at
a central QPC~\cite{ines_PRB_R_noise_2020}.\footnote{A related idea was
developed in Ref.~\cite{fractional_statistics_zhang_Gefen_2025_anyon_collider},
although restricted to a TLL model.}

Building on this result, we introduce two minimal single-QPC strategies
to access $\theta$, based on a \emph{local} and robust anyonic
time-exchange (ATE) relation that yields two fluctuation-dissipation
relations (FDRs), extendable to finite temperatures and to
nonequilibrium three-terminal setups. Their scope and implications are
discussed in the concluding section.

\paragraph{Braiding of anyons in the time domain}
Although braiding one anyon around another cannot be strictly realized
along a one-dimensional edge, signatures of anyonic statistics are
encoded in the bosonized fields. Hierarchical FQH states are therefore
essential, since their statistical phase and fractional charge are not
uniquely fixed by the filling factor, unlike in Laughlin states. We
focus on a single edge of the Hall bar and adopt the Fröhlich
classification~\cite{Frohlich_FQHE_1997,Frohlich_FQHE_2001}, where each $K$ matrix underlies each model candidate. We thus consider a chiral Abelian field
theory with $N$ bosonic fields
$\vec{\phi}=(\phi_1,\dots,\phi_N)$ and characterize an anyon species by
a real vector $\vec{l}$ ("anyon $\vec{l}$"):
\begin{equation}
\label{eq:Psi_phi}
\Psi^{\dagger}(x)=\frac{e^{-ik_Fx}}{\sqrt{2\pi l_c}}e^{i\vec{l}\cdot\vec{\phi}(x)},
\end{equation}
where $l_c$ is a short-distance cutoff and $k_F$ a Fermi wave number.
The anyon carries charge $e^*=e\,\vec{Q}\cdot\vec{l}$, with $\vec{Q}$
encoding the electromagnetic coupling and satisfying
$\nu=|\vec{Q}|^2$.  The charge density operator is
\begin{equation}\label{eq:rho}
\rho(x,t)=e\,\vec{Q}\cdot\partial_x\vec{\phi}(x,t)/2\pi ,
\end{equation}
while the current operator reads
\begin{equation}\label{eq:current}
j(x,t)=e\,\vec{Q}\cdot\vec{j}(x,t),\qquad
\vec{j}(x,t)=-\partial_t\vec{\phi}(x,t)/2\pi .
\end{equation}
Time-domain braiding involving the statistical phase requires all modes
to be co-propagating (see End Matter (EM) \ref{app:counter}).\footnote{This revisits the extension proposed in
\cite{fractional_statistics_theory_2016}, where conclusions derived for
a single-mode edge were suggested to remain valid for
counterpropagating multimode edges, a situation that was not analyzed
there in a fully controlled multimode framework.}with equal-time
commutator ($\delta_{i,j}$ is the Kronecker symbol)
\begin{equation}
\label{eq:commutator}
[\phi_i(x,0),\phi_j(y,0)]
=i\pi\,\delta_{i,j}\,\mathrm{sign}(x-y).
\end{equation}
This leads to the spatial exchange relation
\begin{equation}
\label{eq:braiding_space}
\Psi(x,0)\Psi^{\dagger}(y,0)
=
e^{i\theta\,\mathrm{sign}(y-x)}
\Psi^{\dagger}(y,0)\Psi(x,0),
\end{equation}
with statistical phase
\begin{equation}
\label{eq:phase}
\theta=\pi|\vec{l}|^2 .
\end{equation}
Notice that one has also a local scaling dimension given by $\delta=\theta/\pi$. We assume that a single species dominates, thus has the lowest scaling dimension. For instance, at $\nu=2/5$, one has two possibilities for the matrix $K$ that yields either  $e^*=2e/5$, and $\theta=4\pi/5$ or $e^*=e/5$, $\theta=\pi/5$ \footnote{This classification does not take into account the QPC, which can lead to a charge $e/3$.}.

To describe time-resolved injection at position $x_{\mathrm{inj}}$, we
use the original nonequilibrium bosonization formalism
\cite{ines_schulz,*ines_prb_long,ines_ann,ines_epj}, based on solving
the equations of motion with time-dependent boundary
conditions.\footnote{This formulation allows time-dependent classical
voltages, unlike later approaches
\cite{out_of_equilibrium_bosonisation_eugene_PRL_2009,
out_of_equilibrium_bosonisation_gutman_epl_2010}, which are restricted
to DC voltages but allow non-Gaussian sources.}
Further details for short-range interactions are given in
EM~\ref{app:interactions}. Edge interactions determine the
edge-magnetoplasmon transfer matrix ${\bf C}$ with elements
\begin{equation}
\label{eq:C}
C_{ij}(x,y;t-t')
=
[\phi_i(x,t),\phi_j(y,t')] .
\end{equation}
For a quadratic Hamiltonian these $c$-numbers translate $\vec{\phi}$
(for $x\ge x_{\mathrm{inj}}$) as
\begin{equation}
\label{eq:central_solution}
\vec{\phi}(x,t)
=
i\!\int_{-\infty}^{\infty}\! dt'\,
{\bf C}(x,x_{\mathrm{inj}};t-t')
\cdot\!
\vec{j}(x_{\mathrm{inj}},t').
\end{equation}
Consider a first injection of a single anyon $\vec{l}$ with charge $e^*$
at time $t_{\mathrm{inj}}$,
\begin{equation}\label{eq:jinj_l}
\vec{j}(x_{\mathrm{inj}},t')
=\delta(t'-t_{\mathrm{inj}})\vec{l}.
\end{equation}
Using Eq.~\eqref{eq:commutator}, this produces a density pulse
$\vec{\rho}(x,t_{\mathrm{inj}})=\delta(x-x_{\mathrm{inj}})\vec{l}$  thus
$\rho(x,t_{\mathrm{inj}})=e^*\delta(x-x_{\mathrm{inj}})$. For
$x\ge x_{\mathrm{inj}}$ this translates
$\vec{\phi}(x,t)\!\to\!\vec{\phi}(x,t)+
i{\bf C}(x,x_{\mathrm{inj}};t-t_{\mathrm{inj}})\cdot\vec{l}$,
transforming the anyon field as
\begin{equation}\label{eq:phase_injection}
\Psi^{\dagger}(x,t)\to
\Psi^{\dagger}(x,t)
e^{-\,\vec{l}^{\,T}\cdot{\bf C}(x,x_{\mathrm{inj}};t-t_{\mathrm{inj}})\cdot\vec{l}} .
\end{equation}
Inter-mode interactions render ${\bf C}(x,x_{\mathrm{inj}};t-t_{\mathrm{inj}})$
non-diagonal and nonuniversal, unlike the instantaneous case
$t=t'$ where Eq.~\eqref{eq:commutator} ensures diagonal form. For
short-range interactions the $N$ modes $\phi_i$ transform into proper
modes with velocities $\tilde v_m$, giving
\begin{equation}
\label{eq:phase_C_splitted}
\vec{l}^{\,T}{\bf C}(x,y;t-t')\vec{l}
=
i\pi\sum_{m=1}^N
\delta_m\,
\mathrm{sgn}(x-y-\tilde v_m(t-t')) ,
\end{equation}
with nonuniversal scaling dimensions $\delta_m$. Importantly,
\begin{equation}\label{eq:identity_scaling}
\sum_{m=1}^N\delta_m=\delta=\theta/\pi ,
\end{equation}
where $\delta$ is the local scaling dimension at $x=y$, which is
topologically protected. This parallels the integer quantum Hall case,
where interactions cancel locally at a QPC and yield $\delta=1$.

The universal phase $\theta$ therefore splits into $N$ kinks of heights
$\pi\delta_m$, consistent with fractionalization of the injected charge
$e^*$ into $N$ partial charges
\cite{ines_epj,ines_ann,fraction_edge_states_eugene_PRB_2008}. Each
partial charge carries a non-protected phase $\pi\delta_m$, similarly to
the integer regime~\cite{fractionnalisation_anyon_collider_Mora_2021}.
Recovering $\theta$ thus requires suppressing this spreading, ideally
through local injection at $x_{\mathrm{inj}}=x$. In this limit
Eq.~\eqref{eq:phase_C_splitted} yields $\vec{l}^{\,T}\!\!\cdot{\bf C}(x,x;t-t_{\mathrm{inj}})\cdot\!\vec{l}
=
-i\theta\,\mathrm{sgn}(t-t_{\mathrm{inj}}),$
so that Eq.~\eqref{eq:phase_injection} gives
\begin{equation}\label{eq:Psi_braiding}
\Psi^{\dagger}(x,t)\to
\Psi^{\dagger}(x,t)e^{i\theta\,\mathrm{sgn}(t-t_{\mathrm{inj}})}\end{equation}
This robustness extends to finite-range interactions. Indeed, in
Eq.~\eqref{eq:central_solution} an arbitrary $\vec j(x_{\mathrm{inj}},t')$
must be recovered when $x=x_{\mathrm{inj}}$, implying
\begin{equation}
\label{eq:C_local}
{\bf C}(x,x;t-t')
=
-i\pi\,\mathrm{sign}(t-t')\,{\bf Id}.
\end{equation}
This identity will be central for the analysis of the local ATE link.
In case one has $x_{inj}\neq x$, time-domain braiding between an injected anyon and the anyon-quasihole pair created at the QPC ~\cite{fractional_statistics_theory_Sim_Nat_comm_2016,*Sim_non_abelian_NATcomm_2022,oreg_PRL_FQHE_statistics_2023,fractional_statistics_heiblum_sim_nature_2023,martin_anyons_finite_width_FQHE_PRL_2024,matteo_PRB_anyons_2025,fractional_satistics_zhang_gefen_PRL_2025,feve_anyons_martin_Science_2025} with a phase $\theta$ requires two stringent conditions: \emph{(i)} absence of inter-edge interactions, thus free chiral bare modes $\phi_i(x,t)=\phi_i(x-v_i t,t=0)$, and  \emph{(ii)} equal mode velocities, $v_i=v$. Then one has a diagonal ${\bf C}(x,x_{inj};t-t_{inj})=i\pi\,\text{sign}\left[x-x_{inj}-v\,(t-t_{inj})\right]\bf{Id}$, and Eq.\eqref{eq:braiding_psi_interactions} (see Eq.~\eqref{eq:phase}) simplifies to:
\begin{equation}\label{eq:oreg}
\Psi^{\dagger}(x,t)\longrightarrow
\Psi^{\dagger}(x,t)\,
e^{-i\,\theta\,\text{sign}\left[x-x_{inj}-v(t-t_{inj})\right]}.
\end{equation}
 We notice that the proper modes cannot have equal velocities in presence of inter-edge interactions which do not allow one to reduce Eq.\eqref{eq:phase_C_splitted} to a plasmonic kink with height $\theta$ .

Voltage pulses $V(t)$ can shape the charge content and temporal profile of injected
anyonic wave packets, as originally proposed in
~\cite{ines_schulz,*ines_prb_long,ines_ann,ines_epj} and exploited in recent
time-domain interferometry experiments and proposals
\cite{martin_anyons_dip_HOM_PRL_2023,feve_anyons_martin_Science_2025}.
Two experimental caveats should however be noted. First, nonuniversal
renormalization effects along the propagation path, analogous to those in
Eq.~\eqref{eq:phase_injection}, may hinder a universal identification of
the resulting phase with $\theta$. Second, even if such effects are suppressed, $\theta$ cannot in general be
accessed by a single voltage pulse
$V(t)=h\delta(t-t_{\mathrm{inj}})/e$ coupled equally to all edge
modes~\cite{martin_anyons_dip_HOM_PRL_2023,matteo_PRB_anyons_2025}.
Such a pulse rather produces a phase given by $\pi e^*/e$ (see EM \ref{app:interactions}). 

Let us now consider stationary injection of anyons,
$\vec{j}(x_{\mathrm{inj}},t')=\vec{j}_{\mathrm{inj}}$.
From Eqs.~\eqref{eq:central_solution} and~\eqref{eq:commutator} one finds
the stationary solution
$\partial_t\vec{\phi}(x,t)=-2\pi\vec{j}_{\mathrm{inj}}$,
reflecting the stationarity of the average DC current
$I_{\mathrm{inj}}=e\,\vec{Q}\!\cdot\!\vec{j}_{\mathrm{inj}}$
(see Eq.~\eqref{eq:current}).
Introducing $\dot N_{\mathrm{inj}}=I_{\mathrm{inj}}/e^*$ gives
$\vec{j}_{\mathrm{inj}}=\dot N_{\mathrm{inj}}\,\vec{l}$,
since $e^*=e\,\vec{Q}\!\cdot\!\vec{l}$.
Consequently, the phase $\vec{l}\!\cdot\!\vec{\phi}(x,t)$ in
Eq.~\eqref{eq:Psi_phi} is shifted by $2N_{\mathrm{inj}}\,\theta$.
This provides a rigorous proof that the parameter $\lambda$,
introduced to account for DC renormalization effects in the
``anyon collider''~\cite{fractional_statistics_theory_2016},
is universal, $\lambda=\theta/\pi$ (see Eq.~\eqref{eq:phase}),
without assuming noninteracting modes with equal velocities as in the
phenomenological argument of
\cite{fractional_statistics_theory_Sim_Nat_comm_2016,
*Sim_non_abelian_NATcomm_2022,oreg_PRL_FQHE_statistics_2023}.
This universality is related to that of the quantized DC conductance,
equal to $\nu e^2/h$ in FQH states and to $e^2/h$ in interacting
quantum wires~\cite{ines_schulz,*ines_prb_long,ines_ann,ines_epj,maslov_g}.
The partial charges carried by the $N$ modes sum to $e^*$ and their
phases $\pi\delta_m$ to $\theta$, just as fractionalized charges
transmitted through a wire sum to the injected charge in the DC regime.

However, cross-correlations in the "anyon-collider'' geometry do not
separate the role of $\theta$ as a genuine statistical phase from that as a scaling dimension, and they are further restricted to
zero temperature. A more robust and model-independent approach,
proposed in~\cite{ines_PRB_R_noise_2020}, applies to arbitrary inter-edge
interactions and finite temperature. It relies on determining the
effective voltage drop at the central QPC 
which involves $\sin(2\theta)$ (see Eq.~\eqref{eq:omegadc_colldier}).
The complementary methods proposed below require only a single QPC,
while remaining applicable to the ``anyon collider'' setup.
\paragraph{Anyonic time exchange (ATE) link.} We first write the exchange relation between the anyon fields in Eq.~\eqref{eq:Psi_phi}. 
Remarkably, the Hausdorff-Campbell formula yields exactly the same matrix ${\bf C}$ as in Eq.~\eqref{eq:C},
\begin{equation}
\label{eq:braiding_psi_interactions}
\Psi(x,t)\Psi^{\dagger}(y,t')
= e^{\vec{l}^{T}\!\cdot{\bf C}(x,y;t-t')\cdot\vec{l}}
\,\Psi^{\dagger}(y,t')\Psi(x,t).
\end{equation}
We base our analysis on the \emph{local} time-exchange relation obtained at $x=y$, which is
more protected than its nonlocal counterpart in Eq.~\eqref{eq:phase_injection}, as shown through Eq.\eqref{eq:C_local} used for the local anyon injection.
 One thus obtains the \emph{local}
anyon-quasihole exchange link in the time domain (see Ref.~\cite{ines_guyon} for Laughlin states):
\footnote{The robustness of this local relation against edge reconstruction remains to be assessed.
It is nevertheless expected to persist as long as the bosonic equation of motion is linear and admits
the solution~\eqref{eq:central_solution}.}
\begin{equation}\label{eq:braiding_time}
\Psi(0,t)\Psi^{\dagger}(0,t')
= e^{-i\theta\,\text{sign}(t-t')}\,
\Psi^{\dagger}(0,t')\Psi(0,t).
\end{equation}
Here we fix $x=0$ as the position of the QPC. This local link is in fact
necessary for braiding in the time domain: both schemes involve exactly the same phase, even when it deviates from $\theta$ for counter-propagating modes (see  EM \ref{app:counter}) or possibly due to other nonuniversal features.
\begin{figure}[tbh]
\centering
\includegraphics[width=6cm]{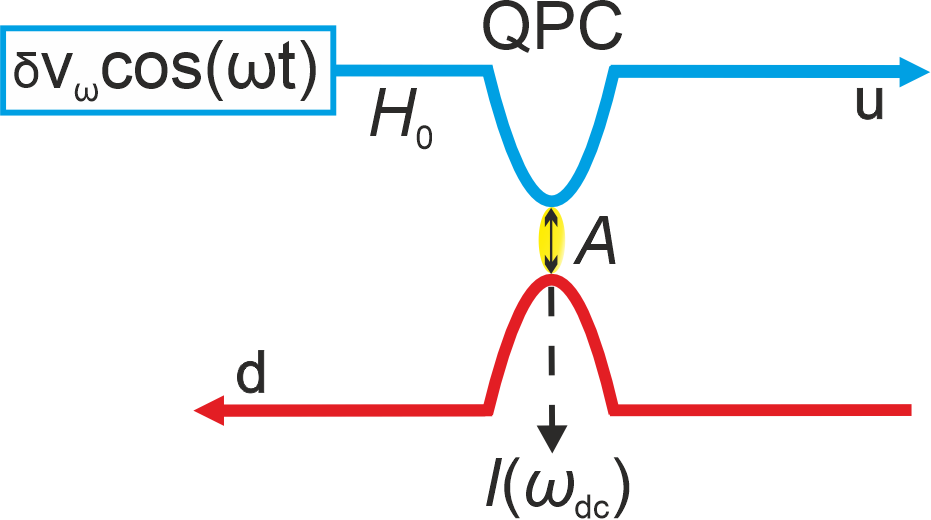}
\caption{A spatially local QPC with a backscattering operator $A$ between the upper (u) and lower edges (d) described by a Hamiltonian $H_0$. The methods for determining $\theta$  apply as well to three terminal setups.}
\label{fig:QPC}
\end{figure}
 
Our next objective is to disentangle the role of $\theta$ as an anyonic
time-exchange (ATE) phase from its appearance as a power-law exponent. To this end,
we apply Eq.~\eqref{eq:braiding_time} to the upper and lower edge fields, labeled
by $u,d$ and carrying opposite chiralities (see Fig.~\ref{fig:QPC}), between which we do not allow for interactions.  Writing the backscattering operator at $x=0$ with amplitude $\xi$, $A(t)=\xi\,\Psi_u^{\dagger}(0,t)\Psi_d(0,t)$,
one immediately infers: $A^{\dagger}(t)A(0)
= e^{-2i{\theta}\,\text{sign}(t)}\,A(0)A^{\dagger}(t)$.
A central idea of this work is to impose this link \emph{on average}
with respect to an arbitrary stationary nonequilibrium distribution
${\rho}_{\mathrm{neq}}$, without specifying the edge Hamiltonian $H_0$.
Our approach can therefore describe \emph{DC} anyon injection, rather than
time-resolved injection, and is directly applicable to DC
``anyon-collider'' setups~\cite{alex_ines_statistics}.
Thus we require that the nonequilibrium correlators
\begin{equation}\label{eq:X}
\hbar^2 X_{\scriptscriptstyle\rm{\rightarrow}}(t)
=\langle A^{\dagger}(t)A(0)\rangle,
\qquad 
\hbar^2 X_{\scriptscriptstyle\rm{\leftarrow}}(t)
=\langle A(0)A^{\dagger}(t)\rangle
\end{equation}
obey the fundamental time-exchange relation which isolates the  role of $\theta$  as an exchange phase from its role as a scaling dimension (see Eq.\eqref{eq:identity_scaling}). \footnote{ $\theta$ intervenes only
modulo $\pi$, as usual.}
\begin{equation}\label{eq:braiding_X}
X_{\scriptscriptstyle\rm{\rightarrow}}(t)
= e^{-2 i\theta\,\text{sign}(t)}\,
X_{\scriptscriptstyle\rm{\leftarrow}}(t).
\end{equation}
\paragraph{ATE nonequilibrium fluctuation-dissipation relation.} Given the Hamiltonian $H_0$ and the backscattering operator $A$, we rely on UNEP theory under three assumptions: $1$- the validity of the second-order perturbation theory with respect to $A$. $2$ -$[{\rho}_{neq},H_0]=0$, ensuring the stationarity of
${\rho}_{neq}$;  $3$- $\langle A(t)A(0)\rangle=0$, a direct consequence of
charge conservation~\cite{ines_PRB_2019}. Then all observables can be expressed in terms of the two
building blocks in Eq.~\eqref{eq:X}. The DC regime is characterized by a drive
frequency $\omega_{\scriptscriptstyle\rm{dc}}$,\footnote{To emphasize its
Josephson-type relation to the DC voltage, $\omega_{\scriptscriptstyle\rm{dc}}$
was denoted $\omega_J$ in Ref.~\cite{ines_PRB_R_noise_2020}.} typically related to a
DC bias $V$ through $\omega_{\scriptscriptstyle\rm{dc}}=e^*V/\hbar$
~\cite{wen_photo_PRB_91,ines_eugene,glattli_photo}. In nonequilibrium situations,
$\omega_{\scriptscriptstyle\rm{dc}}$ may acquire a more involved form
(see Eq.~\eqref{eq:omegadc_colldier} for the `"anyon collider'') and is therefore treated as a free parameter.

The ATE link in Eq.~\eqref{eq:braiding_X} strongly restricts the class of
models covered by UNEP theory~\cite{alex_ines_statistics} and renders
$X_{\scriptscriptstyle\rm{\rightarrow}}(\omega)$ and
$X_{\scriptscriptstyle\rm{\leftarrow}}(\omega)$ interdependent. A single function then
suffices to determine time-dependent transport, chosen as the retarded correlator
\begin{equation}\label{eq:XR}
X^{\scriptscriptstyle\rm{R}}(t)=\Theta(t)\left[
X_{\scriptscriptstyle\rm{\rightarrow}}(t)
- X_{\scriptscriptstyle\rm{\leftarrow}}(t)
\right],
\end{equation}
which yields the DC current expressed as a function of $\omega_{dc}$ instead of a DC voltage~\cite{ines_eugene,ines_PRB_2019}:
\begin{equation}\label{eq:dc_currentXR}
I(\omega_{\mathrm{dc}})=
2e^* \Re\!\left[X^{\scriptscriptstyle\rm{R}}(\omega_{\scriptscriptstyle\rm{dc}})\right].
\end{equation}
Based on Eq.\eqref{eq:braiding_X}, we further show that (see EM~\ref{app:proof})
\begin{align}\label{eq:S_FDT_braiding}
S(\omega_{\mathrm{dc}})
=-2e^{\scriptscriptstyle\rm{*2}}\cot\theta\;
\Im\!\left[X^{\scriptscriptstyle\rm{R}}(\omega_{\mathrm{dc}})\right].
\end{align}
This \emph{ATE nonequilibrium FDR} is the central result of this work that provides multiple routes to extract 
 $\theta$. We can check that it holds for the TLL model in both initially thermalized
two-terminal and "collider geometries" (see Supplementary Material (SM)) with a scaling dimension $\delta={\theta}\, (\text{mod}\pi)/\pi$. In the fermionic limit $\theta=\pi$, the
apparent divergence is compensated by the $\theta$-dependence of
$X^{\scriptscriptstyle\rm{R}}(\omega_{\scriptscriptstyle\rm{dc}})$.  In the following,
we present two methods based solely on DC noise and its comparison with the
average current, either at finite DC drive or at zero DC drive with weak AC phase
modulation.
\paragraph{Method 1.} We now use the Kramers-Kr\"onig relation for $X^{\scriptscriptstyle\rm R}$. To ensure controlled convergence of the integral, we focus on the odd and even components of $I$ and $S$:
\begin{align}
2\bar{I}(\omega_{\scriptscriptstyle\rm{dc}})=&I(\omega_{\scriptscriptstyle\rm{dc}}) -I(-\omega_{\scriptscriptstyle\rm{dc}}), \label{eq:symmetric_current} \\
2\bar{S}(\omega_{\scriptscriptstyle\rm{dc}})=&S(\omega_{\scriptscriptstyle\rm{dc}}) +S(-\omega_{\scriptscriptstyle\rm{dc}}). \label{eq:symmetric_noise}
\end{align}
We get an ATE integral equation based on Eqs.\eqref{eq:dc_currentXR},\eqref{eq:S_FDT_braiding}:
\begin{align}
\bar{S}(\omega_{\scriptscriptstyle\rm{dc}})-\bar{S}(0)
=&{2e^*}\,\cot\theta \; \textit{P}.\!\textit{V}.\!\int_0^{\infty}\! \frac{d\zeta}{\pi}\,
\frac{\bar{I}(\zeta\;\omega_{\scriptscriptstyle\rm{dc}})}{\zeta(1-\zeta^2)}\label{eq:method_1}.
\end{align}
This constitutes an alternative ATE non-equilibrium FDR, providing a method to extract $\theta$ once the fractional charge $e^*$ is known. It requires sufficiently broad DC-drive coverage and suitable high-frequency regularization; yet, the integral is dominated by frequencies near $\omega_{\scriptscriptstyle\rm dc}$ due to the kernel $1/(1-\zeta^2)$. Thus, the relevant frequency window is naturally constrained by the regime where the effective model and perturbation theory apply. In the TLL model,  Eq.~\eqref{eq:method_1} holds only for $\delta={\theta}\, (\text{mod}\pi)/\pi<1$ to ensure convergence, contrary to its more  general form in Eq.\eqref{eq:S_FDT_braiding}; taking $\omega_{\scriptscriptstyle\rm dc}\ll\omega_c$ ensures $\omega'\ll\omega_c$ where $\omega_c$ is the UV cutoff for TLL validity. The fermionic limit $\delta=1$ is also smooth since the $\delta$-dependent integral vanishes at $\delta=1$.

\paragraph{Method 2.}
We now focus on the backscattering admittance at a frequency $\omega$
$Y_{\omega}=G_{\omega}+iB_{\omega}$  \cite{gabelli_06,gabelli_07,admittance_regul_2004,feve_buttiker_admittance_IQHE_PRL_2007,fractionnalsiation_admittance_fujisawa_PRB_2012}, taken at zero DC drive
($\omega_{\scriptscriptstyle\rm dc}=0$) for simplicity. It is defined
through the response of the AC-averaged current to
$\omega\,\delta\varphi(\omega)$, where $\delta\varphi(\omega)$ is a small
phase felt by the QPC. In EM~\ref{app:formal}, we express the susceptance
$B_{\omega}$ in terms only of the nonequilibrium DC noise $S(\omega_{{\scriptscriptstyle\rm dc}}=\omega)$ and the statistical
phase $\theta$ (see Eq.~\eqref{eq:suscpectance_noise}) leading to
 an FDR for the phase $\phi_{\scriptscriptstyle\rm\omega}$ of the admittance \cite{admittance_IQHE_yin_PRA_2019}, thus shift
 of the AC current (see Eq.~\eqref{eq:current_G}), $\tan\phi_{\scriptscriptstyle\rm\omega}={B_{\omega}}/{G_{\omega}}$. For $\omega\geq 0$, one has
$0\leq\phi_{\scriptscriptstyle\rm\omega}\leq\pi$. A key advantage of this ratio is that,
in the perturbative regime, it is independent of the QPC reflection
coefficient $R$. It also cancels any common renormalization of the
voltage and current at the QPC by a purely real or complex factor,
but inter-edge interactions may require a low enough
$\omega$ (see SM) for which the method still applies. The admittance phase is therefore more
robust than the susceptance alone and provides a natural built-in
calibration through the AC conductance. This is particularly valuable
when the precise AC voltage amplitude at the QPC is not known.
The FDR for the admittance phase reads (see
EM~\ref{app:formal})
\begin{equation}
\label{eq:ratio_omegaJ=0}
\tan\phi_{\scriptscriptstyle\rm\omega}
= -\tan\theta \cdot
\frac{\bar{S}(\omega_{\scriptscriptstyle\rm dc}=\omega)
- S(\omega_{\scriptscriptstyle\rm dc}=0)}
{e^*\,\bar{I}(\omega_{\scriptscriptstyle\rm dc}=\omega)} .
\end{equation}
One can therefore get access to
$\theta \;\mathrm{mod}\,\pi$ by comparing the nonequilibrium ratio on the  right-hand side, evaluated at a finite DC
drive $\omega_{\scriptscriptstyle\rm dc}=\omega$, with the
experimentally measured admittance phase $\phi_{\omega}$ at AC
frequency $\omega$ and zero DC drive
($\omega_{\scriptscriptstyle\rm dc}=0$).  For non-equilibrium initial states, such
as unequilibrated edges, temperature gradients, non-thermalized QPC,
or in the presence of QPC sources
\cite{fractional_statistics_gwendal_science_2020,fractional_statistics_heiblum_sim_nature_2023},
the noise is super-Poissonian, so that
$\bar{S}(\omega_{\scriptscriptstyle\rm dc})\geq
e^*|\bar{I}(\omega_{\scriptscriptstyle\rm dc})|$, and $S(0)$ reflect
genuine non-equilibrium fluctuations
\cite{ines_cond_mat,ines_PRB_R_noise_2020}.

\paragraph{Application to thermal states.} We now consider an initially thermalized system at temperature $T$, for
which $S(\omega_{\scriptscriptstyle\rm dc}=0)=2e^*\omega_{\rm th}G(0)$
is the equilibrium noise, with
$G(\omega_{\scriptscriptstyle\rm dc}=0)=G_{\omega=0}$, the linear rescaled conductance, and
$\omega_{\rm th}=k_BT/\hbar$. Assuming that the quantum regime
$\omega\gg\omega_{\scriptscriptstyle\rm th}$ can be reached, and that $S(\omega_{\scriptscriptstyle\rm dc}=0)$ is
negligible with respect to the non-equilibrium noise
$\bar{S}(\omega_{\scriptscriptstyle\rm dc}=\omega)
= e^*\bar{I}(\omega_{\scriptscriptstyle\rm dc}=\omega)$, the FDR in Eq.~\eqref{eq:ratio_omegaJ=0} yields a remarkably simple relation 
\begin{equation}\label{eq:simple}
    \tan\phi_{\scriptscriptstyle\rm\omega}\simeq -\tan\theta ,
\end{equation}
which provides a direct probe of  $\theta$.

Remarkably, we show, in ~\cite{alex_ines_statistics}, that the time-exchange relation
in Eq.~\eqref{eq:braiding_X}, when combined with detailed balance, fully determines a
TLL-like DC current characteristic obeying inversion symmetry, $\bar{I}=I$, with a
scaling dimension $\delta=\theta/\pi$ modulo an integer. In view of our previous analysis, this behavior remains robust even when the Hamiltonian does not reduce to a
chiral TLL, provided spatial locality at the QPC is preserved, also protecting 
$\delta$ against
arbitrary inter-edge interactions. This is the case even for counter propagating modes when $\pi\delta\neq\theta$ (see Eqs.\eqref{eq:theta_counter},\eqref{eq:thetabar_counter}). As a consequence,
constraints associated with the weak-backscattering regime (see
~\cite{ines_imen_2025} and SM) imply that
Eq.~\eqref{eq:simple} holds only for $\delta>1/2$ and a sufficiently weak reflection
coefficient $R$. 
\paragraph{Discussion and conclusion}
Our results clarify the status of time-domain braiding in FQH edge states with a QPC in case the injected anyon is the same species as that dominating backscattering at the QPC.\footnote{After writing this manuscript, we became aware of a work addressing the case of a different species in \cite{fractional_statistics_christian_splatt_FQHE_2026} also based on the original form of nonequilibrium bosonisation \cite{ines_epj,ines_ann}.} Contrary to a widespread implicit assumption,
the phase accumulated by a propagating injected anyon is generically not
the universal statistical angle $\theta$. In interacting hierarchical edges,
charge fractionnalization distributes this phase among several components
determined by the corresponding scaling dimensions, each sensitive to microscopic interaction details.
What remains protected, however, is a strictly local quantity: their sum,
which coincides with the scaling dimension at the QPC and equals
$\theta$ modulo $\pi$.

A single phase kink of height $\theta$, required for genuine time-domain braiding,
appears only under restrictive conditions: noninteracting copropagating modes
with identical velocities, or local injection at the QPC.
Thus locality at the QPC plays a central role:
rather than a technical simplification, it is the mechanism that stabilizes the
statistical phase and enables time-domain anyon braiding.

Even under these conditions, voltage-pulse injection does not
generically yield $\theta$, and similar limitations
may arise in "anyon-collider" setups, possibly accounting for the discrepancies
reported in
.~\cite{fractional_statistics_gwendal_PRX_2023,pierre_anyons_PRX_2023}.
By contrast, using a current-conservation argument in
\emph{stationary} injection schemes, we show that the nonuniversal
renormalization parameter $\lambda$ introduced in
~\cite{fractional_statistics_theory_2016} is, in fact, universal 
$\lambda=\theta/\pi$. Our proof is more general and internally consistent
than previous phenomenological arguments based on the assumption of free modes
with equal velocities
\cite{fractional_statistics_theory_Sim_Nat_comm_2016,
*Sim_non_abelian_NATcomm_2022,oreg_PRL_FQHE_statistics_2023}.

Within the local framework, we identify an anyonic time-exchange (ATE)
link at the QPC as the fundamental structure underlying braiding
phenomena. Unlike propagation-induced dynamical effects, this exchange
structure remains robust against arbitrary edge interactions. It provides  two methods that isolate the role of $\theta$ as an ATE phase: an  integral equation for DC current and noise, and a self-calibrated phase admittance at equilibrium valid at low frequencies and identified with $\theta$ 
in the quantum regime for $\delta>1/2$. We hope this motivates extension of admittance measurements in the integer regime \cite{gabelli_07,admittance_IQHE_yin_PRA_2019,fractionnalsiation_admittance_fujisawa_PRB_2012} to the FQH regime.

More generally, there exists a broader class of models in which the ATE
link holds while $\theta$ is no longer the universal statistical phase,
remaining instead given by $\pi\delta$ but unprotected against
inter-edge interactions, as in the case of interacting counter-propagating
modes. In such situations, the proposed methods 
allow a direct determination of $\delta$ without relying on the scaling
behavior of DC transport. 

  More broadly, our analysis clarifies the hierarchy between propagation,
local exchange, and transport: braiding in time originates from locality,
not from a specific edge model such as a free chiral TLL.

 This provides a
controlled framework for interpreting current and future experiments,
particularly in platforms where interaction effects and edge
reconstruction are minimized
\cite{benjamin_edge_reconstruction_science_2023}. Deviations from the
predicted relations would directly signal a breakdown of locality or of
the underlying exchange structure, thereby offering a stringent test of
the microscopic description of FQH edges. 

Although some aspects of our analysis extend to finite-range interactions
and reconstructed edges, a systematic assessment of the robustness of the
local ATE relation in such situations remains an important direction
for future work. 

\textbf{Acknowledgments.}
This work was supported by the ANR grant "QuSig4QuSense" (ANR-21-CE47-0012). The author thanks Aleksander Latyshev and  Gwendal F\`eve for inspiring discussions, as well as Benoît Douçot, Fr\'ed\'eric Pierre, Gerbold M\'enard, and Benjamin Sac\'ep\'e for valuable suggestions. 
 \bibliographystyle{apsrev4-2}
\begingroup
\small 

\setlength{\abovedisplayskip}{4pt}
\setlength{\belowdisplayskip}{4pt}
\setlength{\abovedisplayshortskip}{2pt}
\setlength{\belowdisplayshortskip}{2pt}

\setlength{\parskip}{0pt}


\setlength{\textfloatsep}{8pt}
\setlength{\floatsep}{6pt}
\setlength{\intextsep}{6pt}

% ...

\endgroup
\appendix
\section*{End Matter}
\section{{Short-range interactions between co-propagating modes}}
\label{app:interactions} 
\paragraph{Phase splitting} Here we evaluate the matrix ${\bf C}$ in
Eq.~\eqref{eq:C} and reconnect with the plasmon-scattering
approach~\cite{ines_schulz,*ines_prb_long,ines_ann,ines_epj,
fraction_edge_states_eugene_PRB_2008,
fractionnalisation_IQHE_degiovanni_13}, which fully encodes inter-mode interactions. For short-range 
interactions between modes on the upper edge, the Hamiltonian reads:
$H_u=\sum_{i,j=1}^N\int dx\, V_{ij}\partial_x\phi_i(x)\partial_x\phi_j(x)$,
without explicitly adding the subscript $u$ to the $N$ bosonic fields
$\phi_i$. 
  We diagonalize $H_u$  through
proper eigenmodes $\Lambda_m$ ($m=1,\dots,N$) propagating freely with
velocity $\tilde v_m$,
$\Lambda_m(x,t)=\Lambda_m(x-\tilde v_m t,0)$. Introducing an
interaction-dependent orthogonal matrix ${\bf O}$ such that
$\vec{\phi}={\bf O}\!\cdot\!\vec{\Lambda}$ yields
\begin{equation}\label{eq:orthogonality}
\sum_{m=1} ^N O_{m,i}O_{m,j}=\delta_{i,j},
\end{equation}
together with the identity $\sum_{m} \!O_{m,i}\tilde v_m O_{m,j}=V_{ij}$.
These relations forbid identical velocities $\tilde v_m$ except in the
case of purely intra-edge interactions.\footnote{I thank Aleksander
Latyshev for drawing my attention to this point for the two-mode case.}

The elements of ${\bf C}$ in Eq.~\eqref{eq:C} then read
\begin{align}\label{eq:C_g_k}
C_{ij}(x,y;t)
&= i\pi\sum_{m=1}^N O_{im}O_{jm}\,g_m(x,y;t),\\
g_m(x,y;t)
&= \mathrm{sign}(x-y-\tilde v_m t).
\end{align}
This yields Eq.~\eqref{eq:phase_C_splitted}, withn
\[
\delta_m=\big[(\vec l^{\,T}\!\!\cdot{\bf O})_m\big]^2 ,
\]
the scaling dimension associated with the combination
$x-y-\tilde v_m t$. 
The consistency of Eqs.~\eqref{eq:central_solution},\eqref{eq:C} with Eq.~\eqref{eq:C_g_k} follows from a direct construction inspired by Refs.~\cite{ines_schulz,*ines_prb_long,ines_ann,ines_epj}: expressing $\vec{\Lambda}(x_{inj},t')={^T\bf O}\!\cdot\vec{\phi}(x_{inj},t')$, propagating freely each eigenmode $\Lambda_k$, and reconstructing $\vec{\phi}(x,t)={\bf O}\!\cdot\!\vec{\Lambda}(x,t)$ reproduces Eq.~\eqref{eq:central_solution} with matrix elements given by Eq.~\eqref{eq:C}.

Using Eq.~\eqref{eq:orthogonality}, we obtain the phase splitting Eq. ~\eqref{eq:phase_C_splitted}.
Thus the phase acquired by "anyon $\vec{l}$" in Eq.\eqref{eq:phase_C_splitted} consists into multiple kinks whose height is given by $\pi\delta_{m}$. 

\paragraph{Local value} At $x=x_{inj}$ one has $g_m(x,x;t)=-\mathrm{sign}(t)$ independently of $\tilde v_m$, so that Eq.\eqref{eq:C_g_k} yields (see Eq.~\eqref{eq:orthogonality})
\[
\vec l^{\,T}\!\cdot{\bf C}(x,x;t)\!\cdot\vec l
=-i\pi\,\mathrm{sign}(t)\sum_{m=1}^N  O_{im}O_{jm}
=-\theta\,\mathrm{sign}(t)\,\delta_{i,j}.
\]
This directly justifies Eqs.~\eqref{eq:C_local} and~\eqref{eq:braiding_time}, and shows that a \emph{single}, robust local scaling dimension $\theta/\pi$ controls the ATE link even in the presence of interactions.

\paragraph{Charge fractionnalisation}
The splitting of $\theta$ into $N$ multiple phases is  related to fractionnalisation of the injected anyon charge $e^*$ into $N $ partial charges $q_m$ freely propagating at velocity $\tilde v_{m}$. In order to express  $q_m$ formally, we  differentiate Eq.\eqref{eq:central_solution}  with respect to $x$, taking into account the injection of anyon  Eq.\eqref{eq:jinj_l}
\begin{equation}
\label{eq:central_solution_density}
\vec{\rho}(x,t)
=
i\partial_x{\bf C}(x,x_{\mathrm{inj}};t-t_{inj})\,
\cdot\vec{l}/2\pi.
\end{equation}
Using Eq.\eqref{eq:C_g_k}  we get
\begin{align}\label{eq:density_short_range} 
\partial_xC_{ij}(x,x_{inj};t)=&2\pi i\sum_{m} O_{i{m}}O_{j{m}}\,\delta(x-x_{inj}-\tilde v_{m}( t-t_{inj})),
\end{align} 
Thus, in view of Eq.\eqref{eq:rho}, the charge fraction per each proper mode reads:\begin{equation}
\label{eq:density_mode}
{q}_{m}
=\sum_{i,j} O_{i{m}}O_{j{m}} Q_i l_j
\end{equation}
so that we have  $\rho(x,t)=\sum_{m} {q}_{m}\delta(x-x_{inj}-\tilde v_{m}( t-t_{inj}))$. Again, using Eq.\eqref{eq:orthogonality}, we get $\sum_m q_m=\vec Q\cdot \vec l=e^*$, in parallel with Eq.\eqref{eq:phase_C_splitted}, which expresses DC conservation of the injected charge.

\paragraph{DC injection} 
Current conservation can be verified explicitly by evaluating the zero-frequency Fourier transform of $\partial_t C_{ij}(x,x_{inj};t)$. For $x\ge x_{inj}$, $\partial_t g_m(x,x_{inj};t)=-2\tilde v_m\,\delta(x-x_{inj}-\tilde v_m t)$, which becomes independent of $m$ after time integration. One obtains, in view of Eq.~\eqref{eq:orthogonality}
\begin{equation}\label{eq:ClocalDC}
\int dt\,\partial_t C_{ij}(x,x_{inj};t)
=-2i\pi\sum_{m=1}^N O_{im}O_{jm}
=-2i\pi\delta_{i,j}.
\end{equation}
Differentiating Eq.~\eqref{eq:central_solution} with respect to time and imposing a constant boundary condition $\vec j_{inj}$ yields $\vec j(x,t)=\vec j_{inj}$ for all $x\ge x_{inj}$. 
\paragraph{Link to the DC conductance} The matrix ${\bf C}$ directly determines the nonlocal conductivity, in close analogy with the scattering approach for interacting quantum wires~\cite{ines_schulz,*ines_prb_long,ines_ann,ines_epj}. The robustness of Eq.\eqref{eq:ClocalDC} guarantees DC conductance quantization despite interactions: $\nu e^2/h$ in the FQH regime and similarily $e^2/h$ in interacting quantum wires. For that, let us recall that voltage couples to the density through $\vec{Q}$. Assuming for simplicity a uniform voltage $V$ applied to all modes, one has $\vec j_{inj}=eV\vec{Q}/h$. This again yields the stationary solution $\vec j(x,t)=eV\vec{Q}/h$. The current in Eq.~\eqref{eq:current} then reads $e\,\vec{Q}\!\cdot\!\vec{j}=e^2\nu/h$, since $\nu=|\vec{Q}|^2$.

\paragraph{Voltage pulse}
For a voltage $V(t)=h\delta(t-t_{inj})$, one has:
$\vec{j}(x_{\mathrm{inj}},t')=\delta(t'-t_{\mathrm{inj}})\vec{Q}$,
which differs from the boundary condition in
Eq.~\eqref{eq:jinj_l} required to inject an anyon $\vec{l}$.
The phase of $\Psi$ in Eq.~\eqref{eq:Psi_phi} is therefore shifted by
the nonuniversal function
$\vec{l}^{\,T}\!\cdot\!{\bf C}(x,x_{\mathrm{inj}};t-t_{\mathrm{inj}})\!\cdot\!\vec{Q}$.
Ensuring either noninteracting modes with equal velocities, or a voltage 
applied locally at the QPC, the phase $\theta$ in Eq.~\eqref{eq:oreg} is
 replaced by $\pi \vec{l}\!\cdot\!\vec{Q}=\pi e^*/e$, determined
by the fractional charge. Only in the non-generic case
$\vec{Q}=\vec{l}$ does this coincide with $\theta$, as in Laughlin states
where $l=Q=\sqrt{\nu}$.
\section{Counterpropagating modes}
\label{app:counter}
Here we address the case where the $N$ modes characterized by $\phi_i$ have chiralities $\sigma_i = \pm 1$ that are not identical. The commutators are then
\begin{equation}
\label{eq:commutator_counter}
[\phi_i(x,0),\phi_{j}(y,0)] = i \sigma_i \pi \, \delta_{i,j} \, \mathrm{sign}(x - y).
\end{equation}
This leads again to Eq.~\eqref{eq:phase} but with the statistical phase for the anyon $\vec{l}$ given by
\begin{equation}
\label{eq:theta_counter}
\theta = \pi \sum_m \sigma_m \, l_m^2.
\end{equation}
There is however no access to $\theta$  through dynamics in time with a single QPC. This can be shown already by assuming free propagation of each mode with velocity $\sigma_iv_i$, we have
$\phi_i(x,t) = \phi_i(x - \sigma_i v_i t, 0)$. The commutators at different times read:
\begin{equation}
\label{eq:C_counter}
C_{ij}(x,y;t-t') = [\phi_i(x,t), \phi_{j}(y,t')] 
= i \pi \, \delta_{i,j} \, \mathrm{sign}\big(\sigma_i (x - y) - (t - t')\big).
\end{equation}
For any anyon injected at position $x_{\mathrm{inj}}$, the absence of a unique chirality must be taken into account. The simplest way to obtain a well-defined phase is local  injection at the QPC, assumed at $x=0$, so that the anyon field at the QPC transforms as in Eq.\eqref{eq:Psi_braiding} with $\theta$ replaced by

\begin{equation}\label{eq:thetabar_counter}
 \bar{\theta} = \pi \sum_m  l_m^2= \pi\delta,
\end{equation}
where $\delta$ is the local scaling dimension. Note that $\bar{\theta}$ differs from $\theta$ in Eq.~\eqref{eq:phase} and is renormalized by inter-edge interactions, contrary to co-propagating edges. This arises because the transformation to proper modes is no longer orthogonal but not detailed here (we refer to \cite{fractional_levkivskyi_PRL_2016} for two modes, Eq.$9b$).

Similarly, the local ATE link involves the same phase $\bar\theta$ as local anyon injection:
\begin{equation}
\label{eq:braiding_space_temporal}
\Psi(x,t) \Psi^{\dagger}(x,t') 
= e^{i \bar{\theta} \, \mathrm{sign}(t' - t)} 
\Psi^{\dagger}(x,t') \Psi(x,t).
\end{equation}
Therefore, for counter-propagating modes, all methods proposed so far to access the braiding phase in time, including the protocols introduced here, probe only the scaling dimension and not the statistical phase governing spatial exchange.

\subsection{Proof of the ATE nonequilibrium FDR}
\label{app:proof}
Within UNEP theory, the DC backscattering current and noise read
\begin{align}
I(\omega_{\mathrm{dc}})&=e^*\!\left[X_{\scriptscriptstyle\rm{\rightarrow}}(\omega_{\scriptscriptstyle\rm{dc}}) 
-X_{\scriptscriptstyle\rm{\leftarrow}}(\omega_{\scriptscriptstyle\rm{dc}})\right],\\
S(\omega_{\scriptscriptstyle\rm{dc}})&=e^{\scriptscriptstyle\rm{*2}}\!
\left[X_{\scriptscriptstyle\rm{\rightarrow}}(\omega_{\scriptscriptstyle\rm{dc}}) 
+X_{\scriptscriptstyle\rm{\leftarrow}}(\omega_{\scriptscriptstyle\rm{dc}})\right],
\end{align}
where the positive $X_{\scriptscriptstyle\rm{\rightarrow}}$ and $X_{\scriptscriptstyle\rm{\leftarrow}}$ are the Fourier transforms of Eq.~\eqref{eq:X}, interpreted as transfer rates~\cite{ines_eugene,*ines_cond_mat} that determine all finite-frequency observables.

Assuming the ATE link in Eq.~\eqref{eq:braiding_X}, we define
$X_{\scriptscriptstyle\rm{\rightarrow}}^{\epsilon}(\omega)=\int dt\,\Theta(\epsilon t)\, e^{i\omega t} X_{\scriptscriptstyle\rm{\rightarrow}}(t)$, $\epsilon=\pm$. Using Eq.~\eqref{eq:braiding_X}, one finds
\begin{align} 
S(\omega_{\scriptscriptstyle\rm{dc}})/e^{\scriptscriptstyle\rm{*2}} 
&=\sum_{\epsilon=\pm}(1+e^{2i\epsilon\theta}) 
X_{\scriptscriptstyle\rm{\rightarrow}}^{\epsilon}(\omega_{\scriptscriptstyle\rm{dc}}), 
\end{align} 
and
\begin{align} 
2i\,\mathrm{Im}\,X^{\scriptscriptstyle\rm R}(\omega_{\scriptscriptstyle\rm{dc}}) 
&=\sum_{\epsilon=\pm}\epsilon(1-e^{2i\epsilon{\theta}}) 
X_{\scriptscriptstyle\rm{\rightarrow}}^{\epsilon}(\omega_{\scriptscriptstyle\rm{dc}}). 
\end{align} 
Eliminating $X_{\scriptscriptstyle\rm{\rightarrow}}^{\epsilon}$ then yields Eq.~\eqref{eq:S_FDT_braiding}.

\section{Admittance within UNEP theory}
\label{app:formal} The driven Hamiltonian reads~\cite{ines_eugene,*ines_cond_mat,ines_PRB_2019,ines_photo_noise_PRB_2022} 
\begin{equation} 
H(t)=H_0+e^{-i\varphi(t)}A+e^{i\varphi(t)}A^{\dagger}, 
\label{eq:hamiltonian} 
\end{equation} 
where $\varphi(t)$ encodes constant or time-dependent drives, not necessarily related to AC voltages, which may be ill-defined in generic nonequilibrium setups (see Fig.~\ref{fig:QPC}). The DC regime corresponds to $\varphi(t)=\omega_{\scriptscriptstyle\rm{dc}}t$. A paradigmatic example is the nonequilibrium "anyon collider" at zero temperature~\cite{ines_PRB_R_noise_2020,alex_ines_statistics}, where the effective DC frequency $\omega_-=\omega_{\scriptscriptstyle\rm{dc}}$, defining a local voltage drop at the central QPC, is related to the injected currents by 
\begin{equation} 
\omega_{\scriptscriptstyle\rm{dc}} 
=\frac{\sin(2\theta)}{e^*}\left(I_{u,inj}-I_{d,inj}\right), 
\label{eq:omegadc_colldier} 
\end{equation} 
with $I_{u,inj}$ and $I_{d,inj}$ the average injected DC currents in the upper and lower edges (see also SM). 

The quasiparticle backscattering current operator is 
\begin{equation}
\hat{I}(t)=\frac{ie^*}{\hbar}\left(e^{-i\varphi(t)}A-e^{i\varphi(t)}A^{\dagger}\right),
\end{equation} 
where $e^*$ denotes the effective quasiparticle charge and may be treated as a free parameter.\footnote{The term ``backscattering'' is used for convenience and does not presuppose a bipartite decomposition of $H_0$.} 

Admittance, depending on both the DC drive frequency $\omega_{\scriptscriptstyle\rm{dc}}$ and the AC frequency $\omega$, can be expressed via exact nonequilibrium FDRs~\cite{ines_proceedings_2009,*ines_philippe} proven useful in a wide range of contexts~\cite{admittance_FF_kondo_andergassen_PRB_13,*admittance_FF_dot_wolfle_PRB_13,*wang_ADN_non_linear_response_2015,*nazarov_back_action_FDR_cite_PRB2019} and recently tested experimentally~\cite{carles_kubo_2025}. A complementary class of exact relations, derived within nonequilibrium bosonized impurity theory, connects the admittance defined via chiral currents to the backscattering admittance, accounting for the renormalization of bosonic fields in Eq.~\eqref{eq:central_solution} together with an analogous renormalization of the applied AC voltage~\cite{ines_bena,*ines_bena_crepieux,*zamoum_12,wang_feldman_FDT_PRB_2011}. Nonuniversal renormalization effects on dephasing are discussed in the SM. Here, we focus exclusively on the \emph{backscattering admittance} at zero DC drive ($\omega_{\scriptscriptstyle\rm{dc}}=0$).
We define its rescaled form under an AC phase $\varphi(t)$,  $Y(t',t)={\delta\langle I_H(t')\rangle}/{\delta\dot{\varphi}(t)}$ where the subscript $H$ denotes the Heisenberg representation with respect to $H(t)$ in Eq.~\eqref{eq:hamiltonian}, and we specialize here to $\omega_{\scriptscriptstyle\rm{dc}}=0$ and a weak AC phase modulation $\varphi(t)=\delta\varphi_{\omega}\sin(\omega t)$. In direct, the AC current response reads 
\begin{align} 
\frac{\delta I_{\omega}(t)}{\omega\,\delta\varphi_{\omega}} 
&= G_{\omega}\cos(\omega t)+B_{\omega}\sin(\omega t)\nonumber\\ 
&= \sqrt{G_{\omega}^2+B_{\omega}^2}\, 
\cos(\omega t+\phi_{\scriptscriptstyle\rm{\omega}}), 
\label{eq:current_G} 
\end{align} 
with $\tan\phi_{\scriptscriptstyle\rm{\omega}}=B_{\omega}/G_{\omega}$. The rescaled admittance is $Y_{\omega}=G_{\omega}+iB_{\omega}$, where $G_{\omega}$ and $B_{\omega}$ are extracted from the in-phase and quadrature components of the current. The ambiguity $\phi_{\scriptscriptstyle\rm{\omega}}\!\!\mod\pi$ is fixed by imposing $\phi_{\omega=0}=0$, since no dephasing is expected in the DC limit. 

Once weak backscattering is addressed within UNEP theory, $Y_{\omega}$ can be expressed to second order in $A$ in terms of the retarded correlator $X^{\scriptscriptstyle\rm{R}}$ (Eq.~\eqref{eq:XR}), 
\begin{equation} 
\hbar\omega Y_{\omega} 
=e^{*2}\Big[-X^{\scriptscriptstyle\rm{R}}(\omega) 
+X^{\scriptscriptstyle\rm{R}*}(-\omega) 
+2i\,\Im X^{\scriptscriptstyle\rm{R}}(0)\Big]. 
\label{eq:admittance_XR} 
\end{equation} 
On one hand, using the DC current expression in Eq.~\eqref{eq:dc_currentXR}, one recovers the UNEP relation for the AC conductance~\cite{ines_eugene,ines_cond_mat,ines_PRB_R_noise_2020}  $\omega G_{\omega}=\bar{I}(\omega_{\scriptscriptstyle\rm{dc}}=\omega)$
where $\bar{I}$ is the odd part of the backscattering current (see Eq.\eqref{eq:symmetric_current}). On the other hand, combining Eqs.~\eqref{eq:S_FDT_braiding}, \eqref{eq:admittance_XR}, and \eqref{eq:symmetric_noise}, we obtain the susceptance: 
\begin{equation} 
e^*\omega B_{\omega} 
=-\tan{\theta}\, 
\Big[\bar{S}(\omega_{\scriptscriptstyle\rm{dc}}=\omega) 
-S(\omega_{\scriptscriptstyle\rm{dc}}=0)\Big]. 
\label{eq:suscpectance_noise} 
\end{equation} 
Thus the admittance at finite frequency $\omega$ and zero DC drive is directly related to DC current and noise evaluated at $\omega_{\scriptscriptstyle\rm{dc}}=\omega$. The ratio $B_{\omega}/G_{\omega}$ provides the FDR for dephasing in Eq.\eqref{eq:ratio_omegaJ=0}. Since $G_{\omega}$ and $B_{\omega}$ obey Kramers-Kronig relations, we recover Eq.\eqref{eq:method_1}.

\section{Supplemental Material: Robust protocols to reveal anyonic time-exchange phase}

\subsection{Starting with a TLL-type behavior}
Here, for TLL type predictions both in a two-terminal thermalized setup at arbitrary temperatures and in an "anyon collider"., we check the fundamental ATE relation for DC noise \begin{align}\label{eq:S_FDT_braiding_supp}
S(\omega_{\mathrm{dc}})
=-2e^{\scriptscriptstyle\rm{*2}}\cot\theta\;
\Im\!\left[X^{\scriptscriptstyle\rm{R}}(\omega_{\mathrm{dc}})\right].
\end{align} in terms of the retarded correlator $X^R$, 
\begin{equation}\label{eq:XR_supp}
X^{\scriptscriptstyle\rm{R}}(t)=\Theta(t)\left[
X_{\scriptscriptstyle\rm{\rightarrow}}(t)
- X_{\scriptscriptstyle\rm{\leftarrow}}(t)
\right],
\end{equation}
which also yields the DC current ~\cite{ines_eugene,ines_PRB_2019}:
\begin{equation}\label{eq:dc_currentXR_supp}
I(\omega_{\mathrm{dc}})=
2e^* \Re\!\left[X^{\scriptscriptstyle\rm{R}}(\omega_{\scriptscriptstyle\rm{dc}})\right].
\end{equation}
First, in a two-terminal setup, one can compute $X^R$ in terms of the reflection coefficient $R$ and the scaling dimension $\delta$:\begin{equation}
X^R(\omega)=-2 iR\left[(2\pi)^2\frac{\omega_{\scriptscriptstyle\rm{th}}}{\omega_c}\right]^{2(\delta-1)}\sin(\pi\delta)B(\delta-i\bar{\omega},1-2\delta),
    \end{equation}
    where $B(x,y)=\Gamma(x)\Gamma(y)/\Gamma(x+y)$,  where $\Gamma$ denotes the Gamma function, is the Beta function, and
\[
\omega_{\scriptscriptstyle\rm{th}}=\frac{k_B T}{\hbar},
\qquad
\bar{\omega}=\frac{\omega}{\omega_{\scriptscriptstyle\rm{th}}}.
\]The TLL DC current expression\cite{wen_photo_PRB_91} in Eq.\eqref{eq:dc_currentXR_supp}, 
\begin{equation}
\label{eq:I_TLL}
I(\omega_{\scriptscriptstyle\rm{dc}}=\omega)
=
2\,\omega_{\scriptscriptstyle\rm{th}}\,G(0)\;
\sinh\!\left(\frac{\bar{\omega}}{2}\right)QPC
\frac{\left|\Gamma\!\left(\delta+i\frac{\bar{\omega}}{2\pi}\right)\right|^2}
{\Gamma(\delta)^2},
\end{equation}
\begin{equation}
\label{eq:GT}
G(0)=
\frac{e^{\scriptscriptstyle\rm{*}} R}{2\pi}
\frac{\Gamma(\delta)^2}{\Gamma(2\delta)}
\left[(2\pi)^2\frac{\omega_{\scriptscriptstyle\rm{th}}}{\omega_c}\right]^{2(\delta-1)},
\end{equation}
One can now check that evaluating Eq.\eqref{eq:XR_supp} with $\theta=\pi \, mod \pi$ gives precisely the poissonian noise,  $S(\omega_{\mathrm{dc}})
=-2e^{\scriptscriptstyle\rm{*2}}\cot\theta\;
\Im\!\left[X^{\scriptscriptstyle\rm{R}}(\omega_{\mathrm{dc}})\right]=e^*\coth(\bar{\omega}/2)I(\omega_{\scriptscriptstyle\rm{dc}})$.

Secondly, we consider the "anyon collider" where $I_{u,inj},I_{d,inj}$ are the injected curents in the upper and lower edges. We restrict here to the zero-temperature backscattering current and noise derived in \cite{fractional_statistics_theory_2016}, referring to \cite{alex_ines_statistics} for the finite temperature case:
\begin{eqnarray}
  I(\omega_+,\omega_-)&=&C'{R}\sin \theta \;\Im(\omega_++i\omega_-)^{2\delta-1},\nonumber\\
S(\omega_+,\omega_-)&=&C'e^*{R}\cos\theta\;\Re(\omega_++i\omega_-)^{2\delta-1},\label{eq:anyon_collider_expressions}
\end{eqnarray}
where $C'$ is a non-universal constant  $\omega_+=(\sin\theta)^2 (I_{u,inj}+I_{d,inj})/e^*$, while $\omega_-=\omega_{{\scriptscriptstyle\rm{dc}}}$ is given by
\cite{fractional_statistics_theory_2016,ines_PRB_R_noise_2020}:  
\begin{equation}
\label{eq:omegaJ_anyon} \omega_- = \sin(2\theta)(I_{u,inj}-I_{d,inj})/{e^*}.
\end{equation}
In the expressions for $I,S$ we have replaced $\theta=\pi\delta \, mod (\pi)$ in the trigonometirc functions. 
The retarded correlator now reads:
\begin{equation}
    X^R(\omega_+,\omega)=i\sin\theta \,(-i\omega+\omega_+)^{2\delta-1}.
\end{equation}
One can easily check Eqs.\eqref{eq:S_FDT_braiding_supp},\eqref{eq:dc_currentXR_supp}.

\subsection{Case of detailed balance equation: ending up with TLL-type scaling}
\label{app:TLL} 
Here we address the case of an initial thermalized distribution, thus replacing $\hat{\rho}_{\mathrm{neq}} \rightarrow \hat{\rho}_{\mathrm{eq}} \propto e^{-\beta \mathcal{H}_0}$ with $\beta = 1/T$. Then one has the detailed-balance condition between the two "transfer" rates:
\[
X_{\scriptscriptstyle\rm{\rightarrow}}(\omega) = e^{\beta \omega}\,
X_{\scriptscriptstyle\rm{\leftarrow}}(\omega).
\]
We recall the anyonic time exchange (ATE):
\begin{equation}\label{eq:braiding_X_supp}
X_{\scriptscriptstyle\rm{\rightarrow}}(t)
= e^{-2 i\theta\,\text{sign}(t)}\,
X_{\scriptscriptstyle\rm{\leftarrow}}(t).
\end{equation}
Since the DC noise becomes Poissonian, the expression for the admittance phase, arising from these two links, becomes 
\begin{equation}\label{eq:ratio_omegaJ=0_thermal} 
\tan \phi_{\scriptscriptstyle\rm{\omega}} 
= -\tan\theta \left[\coth({\omega}/2\omega_{th}) 
- \frac{2\omega_{\scriptscriptstyle\rm{th}}\,G(\omega_{\scriptscriptstyle\rm{dc}}=0)} 
{I(\omega_{\scriptscriptstyle\rm{dc}}=\omega)} \right]. 
\end{equation}
Therefore, for any pair $(\omega,\omega_{\scriptscriptstyle\rm{th}})$ within the weak backscattering regime, one could extract $\theta\;(\mathrm{mod}\,\pi)$ by measuring $\phi_{\scriptscriptstyle\rm{\omega}}$, $G(0)$, and $I(\omega_{\scriptscriptstyle\rm{dc}}=\omega)$, without requiring any direct measurement of DC noise. Equation~\eqref{eq:ratio_omegaJ=0_thermal} remains valid down to arbitrarily low frequencies.

It turns out, according to our analysis in \cite{alex_ines_statistics}, that a remarkable feature arises: combining the two links yields a unique TLL solution for the DC backscattering current average given by Eq.\eqref{eq:I_TLL}..
This solution relies on spatial locality at the QPC. It is valid in parallel with the ATE link, shown in the letter to be robust even in presence of interedge interactions, thus when the edge Hamiltonian $H_0$ does not reduce to a chiral TLL.

Notice that even though we have not initially imposed inversion symmetry, it arises from the solution of the ATE integral equation for the DC current (as the DC noise is now poissonnian) which is necessarily odd, $\bar{I}=I$.
Though one cannot write the ATE integral equation at $\delta>1$, as it would be divergent, the TLL-type solution can be extended from $\delta<1$ to $\delta>1$ by analytical continuation \cite{ines_alex_2025}. We have also shown in the previous section that the basic nonequilibrium FDRs for DC current and noise, which are at the origin of this integral equation and therefore have a more extended validity domain, hold regardless of what $\delta$ is. 

Injecting the expression of the DC current in Eq.\eqref{eq:I_TLL} in Eq.\eqref{eq:ratio_omegaJ=0_thermal} yields the explicit one for the admittance phase:
\begin{equation}
\label{eq:ratio_omegaJ=0_TLL}
\tan \phi_{\scriptscriptstyle\rm{\omega}}
=
-\frac{\tan\theta}{\sinh (\bar{\omega}/2)}
\left[
\cosh(\bar{\omega}/2)
-
\frac{\Gamma(\delta)^2}
{\left|\Gamma\!\left(\delta+i\frac{\bar{\omega}}{2\pi}\right)\right|^2}
\right].
\end{equation}
Although $\tan\theta=\tan(\pi\delta)$, we keep track of the role of $\theta$ in the time-domain exchange link.

Consequently, the AC phase shift $\phi_{\scriptscriptstyle\rm{\omega}}$ obeys the same scaling structure, and its frequency dependence can be inferred from its temperature dependence. Since $\omega$ is typically fixed experimentally, a more robust validation consists into plotting both sides of Eq.~\eqref{eq:ratio_omegaJ=0_thermal} as functions of temperature (see Fig.\ref{Fig:phiT01}) .  
\begin{figure}[htb]
 \centering
\includegraphics[width=7cm,clip]{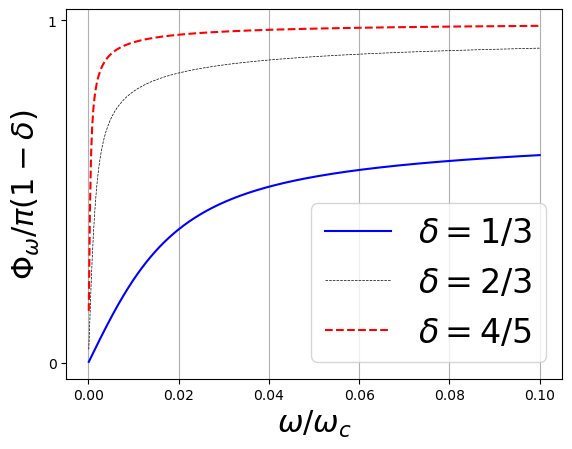}
  \caption{Phase shift for an initial thermalized system from Eq.~\eqref{eq:ratio_omegaJ=0_thermal} f$R=0.01$,  for three values of $\delta=\theta (\text{mod} \;\pi)/\pi$ with their associated lower bounds on $\omega_{th}$ (see section \ref{app:limitation}) given by $4. 10^{-4}$, $ 8.10^{-3}$, $10^{-4}$ $\omega_c$ .}  \label{fig:tanphiomega}
\end{figure}

It is nevertheless instructive to analyze the quantum regime $\omega\gg\omega_{\scriptscriptstyle\rm{th}}$, whose accessibility is discussed in the next section. Once assumed, one can use the power-law dependence of the nonequilibrium DC current on $\omega$ and that of the equilibrium conductance on $\omega_{\scriptscriptstyle\rm{th}}$, so that Eq.~\eqref{eq:ratio_omegaJ=0_thermal} reduces to 
\begin{equation}\label{eq:ratio_scaling} 
\tan \phi_{\scriptscriptstyle\rm{\omega}} 
= -\tan\theta \left[ 
1-\frac{\Gamma(\delta)^2}{\pi} 
\left(\frac{\omega}{2\pi\omega_{\scriptscriptstyle\rm{th}}}\right)^{1-2\delta} 
\right]. 
\end{equation} 
This expression reveals a qualitative distinction between the regimes $\delta<1/2$ and $\delta>1/2$. For $\delta<1/2$, the second term dominates and cannot be neglected, making the full braiding FDR in Eq.~\eqref{eq:ratio_omegaJ=0_thermal} essential. By contrast, for $\delta>1/2$, the first term dominates, with the second contributing only a small correction. In this case, one finds the simple and striking identity:
\begin{equation}\label{eq:simple_SM} 
\tan \phi_{\scriptscriptstyle\rm{\omega}} 
= -\tan\theta.
\end{equation} 
In fact, for $\delta>1/2$, the quantum regime is more accessible as will be discussed in the next section. The explicit frequency dependence of ${\phi_{\scriptscriptstyle\rm{\omega}}}$ from
Eq.~\eqref{eq:ratio_omegaJ=0_thermal} is shown in Fig.~\eqref{fig:tanphiomega} for three values of
$\delta<1$. They demonstrate that the identity  is recovered at high
frequencies for $\delta=2/3$ and $4/5$, but not for $\delta=1/3$.

Whenever one knows in advance whether $\delta<1$, which can be inferred from the qualitative DC voltage or temperature variations of the conductance, it is possible  to infer $\delta$ without ambiguity.  Indeed, for $\delta=1$ the backscattering current is linear, $\langle I(t)\rangle=G(0)\,\dot{\varphi}(t)$, which implies $\phi_{\scriptscriptstyle\rm{\omega}}=0$. Recall that $0\leq\phi_{\scriptscriptstyle\rm{\omega}}\leq\pi$. Requiring its continuity as $\delta\to1$ yields $\delta = 1 - {\phi_{\scriptscriptstyle\rm{\omega}}}/{\pi}$  for $\delta<1$.

\subsection{Limitations on the quantum regime}
\label{app:limitation}
Here we discuss the conditions to ensure the quantum regime, which are similar to those discussed in ~\cite{ines_photo_noise_PRB_2022,ines_imen_2025} for photo-assisted transport . While one might expect to reach this regime by lowering the temperature so that the equilibrium noise $S(0)=2e^*\omega_{\scriptscriptstyle\rm{th}}G(0)$ in Eq.\eqref{eq:ratio_omegaJ=0_thermal} vanishes, this procedure drives the system into the insulating regime, thus requiring instead a finite minimum temperature.

Let us first recall that the transmitted DC current is given by
$G_0\,\omega_{\scriptscriptstyle\rm{dc}}-I(\omega_{\scriptscriptstyle\rm{dc}})$,
where the rescaled perfect conductance is
$G_0=\nu e^2/(2\pi e^*)$, so that
$G_0\,\omega_{\scriptscriptstyle\rm{dc}}=\nu e^2 V/h$.
Let us notice that a counterintuitive feature appears for $\delta<1/2$, where the nonequilibrium conductance
$G(\omega_{\scriptscriptstyle\rm{dc}}\gg\omega_{\scriptscriptstyle\rm{th}})<0$,
meaning that backscattering increases the conductance, while by contrast $G(0)>0$.
Importantly, the weak-backscattering regime imposes that
$G(0)\propto R\,T^{2(\delta-1)}\ll G_0$.
More precisely, following Ref.~\cite{ines_imen_2025}, we require
$G(0)\leq 0.1\,\nu e^2/h$,
which yields an infrared bound
$\omega_{\scriptscriptstyle\rm{th}}\geq \omega_{\scriptscriptstyle\rm{min}}(\delta)$.
Using Eq.~\eqref{eq:GT}, one finds
\begin{eqnarray}
\frac{\omega_{\scriptscriptstyle\rm{min}}(\delta)}{\omega_c}
&=&
\frac{1}{(2\pi)^2}
\left[
\frac{10\,e^{\scriptscriptstyle\rm{*2}}
\Gamma(\delta)^2 R}
{\Gamma(2\delta)\,\nu e^2}
\right]^{\frac{1}{2(1-\delta)}},
\label{eq:minimum_T}
\end{eqnarray}
which decreases with decreasing $R$ and increasing the scaling dimension $\delta$.
Table~\ref{table:cutoffs} gives numerical estimates for $\nu=1/3$,
$e^*=e/3$, and representative values of $R$ and $\delta$.
\begin{table}[tbh]
\begin{center}
\begin{tabular}{|c|c|c|}
\hline
$R$ & $\delta=1/3$ & $\delta=2/3$ \\
\hline
0.1  & $6\times 10^{-2}$ & $1.2\times 10^{-2}$ \\
\hline
0.01 & $8\times 10^{-3}$ & $4\times 10^{-4}$ \\
\hline
\end{tabular}
\end{center}
\caption{Infrared bound in units of the cutoff frequency,
$\omega_{\scriptscriptstyle\rm{min}}(\delta)/\omega_{\scriptscriptstyle\rm{c}}$.}
\label{table:cutoffs}
\end{table}

The quantum regime therefore requires
\begin{equation}
\label{eq:quantum_regime}
\omega_{\scriptscriptstyle\rm{min}}(\delta)
\leq \omega_{\scriptscriptstyle\rm{th}}
\ll \omega
\ll v/L,
\end{equation}
where the additional cutoff $v/L$ avoids possible renormalization due to plasmon
propagation over a distance $L$ (assuming $v/L\ll\omega_c$; see section \ref{app:renormalisation}).
For $\nu=\delta=1/3$ and $R=0.1$, this window disappears, while it remains accessible
for $R=0.01$. In that case, however, the equilibrium noise exceeds the nonequilibrium
noise evaluated at $\omega_{\scriptscriptstyle\rm{dc}}=\omega\gg\omega_{\scriptscriptstyle\rm{th}}$,
so that the second term on the right-hand side of
Eq.~\eqref{eq:ratio_omegaJ=0_TLL} cannot be neglected.

\begin{figure}[thb]
\centering
    \includegraphics[width=8cm,clip]{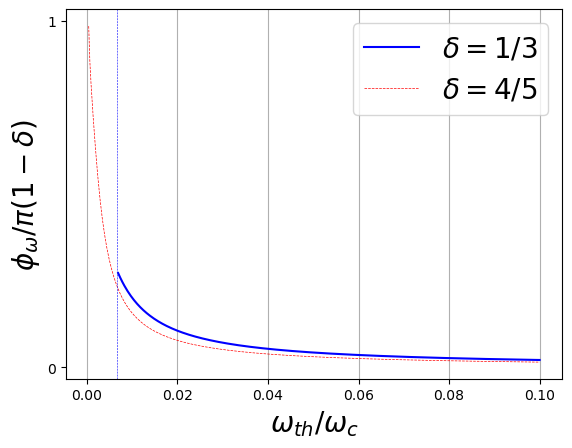}
\caption{TLL model: $T$ or $\omega_{\scriptscriptstyle\rm{th}}$-dependence of $\phi_{\scriptscriptstyle\rm{\omega}}/\pi (-\delta+1)$ at a fixed $ \omega=0.1\;\omega_c$ and two values f $\delta<1$. The minimum value is given by $\omega_{\scriptscriptstyle\rm{min}}({\delta})$,  extremely small for $\delta=4/5$ and indicated by the vertical dotted line for $\delta=1/3$. }\label{Fig:phiT01}
\end{figure}

The temperature dependence of the phase shift $\phi_\omega$ at fixed frequency
is shown in Fig.~\ref{Fig:phiT01}, with lower bounds on temperatures depend on $\delta$ . We see again that the simple identity
$\phi_\omega=\pi(1-\delta)$
is recovered only at very low temperatures and for $\delta>1/2$.

Let us now estimate the ultraviolet cutoff $\omega_c$, assumed to lie below the gap
to bulk quasiparticle excitations. Taking
$a\simeq l_B\simeq 10\,\mathrm{nm}$ for $B=6$--$10\,\mathrm{T}$ and
$v\simeq 10^4$--$10^5\,\mathrm{m/s}$, one finds
$\omega_c\simeq 160$--$1600\,\mathrm{GHz}$.
A smaller effective UV cutoff may arise if the AC phase is imposed through locally
equilibrated left and right reservoirs with electrochemical potentials
$\mu_{L,R}(\omega)$. In that case, inelastic processes with characteristic time
$\tau_{\mathrm{in}}$ must efficiently establish these potentials, which may require
$\omega\tau_{\mathrm{in}}\ll 1$~\cite{ines_epj}.

\subsection{Renormalization effects}
\label{app:renormalisation}
We now discuss the robustness of the phase shift against non-universal renormalization
effects. If such effects originate from a time-domain convolution, they can be modeled
by a frequency-dependent multiplicative factor $r_\omega$ acting on $Y_\omega$.
If $r_\omega$ is purely real or purely imaginary, it cancels out in the phase of the admittance, which involves the ratio between the imaginary and real parts of the admittance (suscpectance over AC conductance)  $\tan\phi_\omega=B_\omega/G_\omega$. This cancellation fails, however, if
$r_\omega$ is complex. This situation may arise, for instance, when the relevant AC
drive is the local phase felt by the QPC, which can differ from the injected tone at
high frequencies due to impedance mismatch or parasitic elements. In such cases,
on-chip calibration may be required, for example using a reference junction or an
in-situ measurement of $Y_\omega$.

The measured current may also be affected by propagation along the chiral edge.
If this propagation is not properly accounted for, the cancellation may again fail,
as in measurements of chiral currents (the ``chiral admittance'') where plasmonic
modes propagate from the QPC over a distance $L$ to the detection points.

For Laughlin states at simple filling factors $\nu$, the nonequilibrium transport
theory for bosonized impurity models~\cite{ines_bena,*ines_bena_crepieux,*zamoum_12}
yields the renormalization factor
\begin{equation}
r_\omega = e^{2 i L \omega / v},
\end{equation}
which accounts equally for the propagation of the AC signal from the left reservoir
to the QPC and for that of the current toward the right reservoir. Each propagation
factor is the Fourier transform of
$\partial_t C(0,L,t)=\delta(L-vt)$, as the matrix $\mathbf{C}$ reduces to a single
function $C$ in the case of a single mode. To avoid oscillatory corrections in
$\phi_{\scriptscriptstyle\rm{\omega}}$, one must therefore restrict the frequencies to
\begin{equation}
\omega \ll v/L.
\end{equation}
In this regime, the quantum condition $\omega_{\scriptscriptstyle\rm{th}}\ll\omega$
also guarantees thermal coherence along the propagation path, since
$\omega_{\scriptscriptstyle\rm{th}}\ll v/L$.

For more complex filling factors $\nu_C$, a more intricate form of $r_\omega$ is
expected, arising from coupling between several edge modes or from edge
reconstruction. Nevertheless, finite-frequency measurements reported in
Ref.~\cite{ines_gwendal} at $\nu_C=2/3$ were successfully analyzed without invoking
oscillatory renormalization factors, suggesting that these effects are either weak
or effectively averaged out under experimental conditions.

\subsection{AC current in the quantum regime}

In the quantum regime one finds that the amplitude of the AC current satisfies
\begin{equation}
\label{eq:main_current}
\frac{\delta I_\omega(t)}{\omega\,\delta\varphi_\omega}
=
\left.
\frac{\bar{I}(\omega_{\scriptscriptstyle\rm{dc}})}
{\omega_{\scriptscriptstyle\rm{dc}}}
\right|_{\omega_{\scriptscriptstyle\rm{dc}}=\omega}
\frac{\cos(\omega t+\theta)}{|\cos\theta|}.
\end{equation}
The response is therefore governed both by the braiding phase $\theta$ and by the
odd part of the DC current evaluated at an effective nonequilibrium voltage, as
implied by Eq.~\eqref{eq:quantum_regime}. In principle, $\theta$ could be extracted
from the prefactor $|\cos\theta|^{-1}$. In practice, this approach is less robust
than using dephasing, since the precise amplitude of the AC driving phase at the
level of the QPC is not known accurately, especially in the quantum regime where
its renormalization may be more pronounced.

\subsection{Case of inversion symmetry without detailed balance}
\label{app:braiding}

We finally consider nonequilibrium distributions obeying the exchange-time link
between anyons and quasiholes under inversion symmetry,
$X_{\scriptscriptstyle\rm{\rightarrow}}(t)
=
X_{\scriptscriptstyle\rm{\leftarrow}}(-t)= X(t)$
or equivalently
$X_{\scriptscriptstyle\rm{\rightarrow}}(\omega_{\scriptscriptstyle\rm{dc}})
=
X_{\scriptscriptstyle\rm{\leftarrow}}(-\omega_{\scriptscriptstyle\rm{dc}})$.
In this case, the current $I$ is odd and the noise $S$ is even in
$\omega_{\scriptscriptstyle\rm{dc}}$.
A single function $X$ therefore enters the time-domain exchange link.
Writing $X(t)=e^{-g(z)}$ with $z=\tau_0+i t$, the most general solution of
Eq.~\eqref{eq:braiding_X} is
$g(z)=(\theta/\pi)\,\log(z/\tau_0)+f(z)$,
where $f$ is analytic separately in the half-planes $\mathrm{Re}\,z>0$ and
$\mathrm{Re}\,z<0$.
This construction also applies to an initial thermal distribution, where the
additional detailed-balance condition leads to
$g(z)=(\theta/\pi)\,\log\!\big[\sin(\pi T z)\big]+f(z)$.
Within the TLL description of the edges, the ``anyon collider'' satisfies this
structure, as will be discussed in a separate publication~\cite{alex_ines_statistics}.
Identifying alternative Hamiltonians $\mathcal{H}_0$ and operators $A$ that lead
to such correlators remains an open problem.

\end{document}